\documentclass[aps,prd,onecolumn,showpacs,groupedaddress,nofootinbib,amssymb]{revtex4}
\usepackage{graphicx}
\usepackage[english]{babel}

\begin{document}

\title{ Higher-order gravity and the cosmological background of gravitational waves}

\author{Salvatore Capozziello$^1$, Mariafelicia De Laurentis$^2$, Mauro Francaviglia$^3$}

\affiliation{\it $^1$Dipartimento di Scienze fisiche, Universit\`a
 di Napoli {}`` Federico II'', and INFN Sez. di Napoli, Compl.
Univ. di
Monte S. Angelo, Edificio G, Via Cinthia, I-80126, Napoli, Italy\\
$^2$Dipartimento di Fisica, Politecnico di Torino, and INFN Sez.
di
Torino, Corso Duca degli Abruzzi 24, I-10129 Torino, Italy\\
$^3$Dipartimento di Matematica, Universit\`a di Torino, and INFN
Sez. di Torino,Via Carlo Alberto 10, 10123 Torino, Italy}

\begin{abstract}
The cosmological background of gravitational waves can be tuned by
the higher-order corrections to the gravitational Lagrangian. In
particular, it can be shown that  assuming $R^{1+\epsilon}$, where
$\epsilon$ indicates a generic (eventually small) correction to
the Hilbert-Einstein action in the Ricci scalar $R$,  gives a
parametric approach to control the evolution and the production
mechanism of gravitational waves in the early Universe.
\end{abstract}

\pacs{98.80.-k, 04.90.+e, 04.50.+h}

\maketitle

{\it Keywords}: extended theories of  gravity; gravitational
waves; cosmology.

\vspace{5.mm}

Several issues coming  from Cosmology and Quantum Field Theory
suggest to extend the Einstein General Relativity in order to cure
shortcomings emerging from observations and self-consistent
unification theories. In early time Cosmology, the presence of Big
Bang singularity, flatness and horizon problems \cite{guth} led to
the result that Standard Cosmological Model \cite{weinberg}, is
inadequate to describe the Universe at extreme regimes. On the
other hand, General Relativity is a \textit{classical} theory
which does not work as a fundamental theory, when one wants to
achieve a full quantum description of spacetime (and then of
gravity). Due to these facts and, first of all, to the lack of a
self-consistent Quantum Gravity theory, alternative theories of
gravity have been pursued in order to attempt, at least, a
semi-classical scheme where General Relativity and its positive
results could be recovered. A fruitful approach has been that of
\textit{Extended Theories of Gravity} which have become a sort of
paradigm in the study of gravitational interaction based on
corrections and enlargements of the Einstein scheme
\cite{odintsov,farhoudi}.

Besides fundamental physics motivations,  these theories have
acquired a huge interest in cosmology due to the fact that they
``naturally" exhibit inflationary behaviors able to overcome the
shortcomings of Standard Cosmological Model (based on General
Relativity). The related cosmological models seem very realistic
and, several times, capable of matching with the observations
\cite{la,kerner}. Recently, Extended Theories of Gravity are going
to play an interesting role to describe the today observed
Universe. In fact, the amount of good quality data of last decade
has made it possible to shed new light on the effective picture of
the Universe. Type Ia Supernovae (SNeIa) \cite{SNeIa},
anisotropies in the cosmic microwave background radiation (CMBR)
\cite{CMBR}, and matter power spectrum inferred from large galaxy
surveys \cite{LSS} represent the strongest evidences for a radical
revision of the Cosmological Standard Model also at recent epochs.
In particular, the \textit{concordance $\Lambda$CDM model}
predicts that baryons contribute only for $\sim4\%$ of the total
matter\,-\,energy budget, while the exotic \textit{cold dark
matter} (CDM) represents the bulk of the matter content
($\sim25\%$) and the cosmological constant $\Lambda$ plays the
role of the so called "dark energy" ($\sim70\%$) \cite{triangle}.
Although being the best fit to a wide range of data
\cite{LambdaTest}, the $\Lambda$CDM model is severely affected by
strong theoretical shortcomings \cite{LambdaRev} that have
motivated the search for alternative models \cite{PR03}. Dark
energy models mainly rely on the implicit assumption that
Einstein's General Relativity is the correct theory of gravity
indeed. Nevertheless, its validity on the larger astrophysical and
cosmological scales has never been tested \cite{will}, and it is
therefore conceivable that both cosmic speed up and dark matter
represent signals of a breakdown in our understanding of
gravitation law so that one should consider the possibility that
the Hilbert\,-\,Einstein Lagrangian, linear in the Ricci scalar
$R$, should be generalized
\cite{frafe,cap,odintsov2,carroll,allemandi}.

From a  genuine astrophysical viewpoint, Extended Theories of
Gravity do not urgently require  to find out candidates for dark
energy and dark matter at  fundamental level (till now they have
not been found!); the approach starts from taking into account
only the ``observed'' ingredients (i.e. gravity, radiation and
baryonic matter); it is in full agreement with the early spirit of
General Relativity which could not act in the same way at all
scales (for comprehensive reviews on the argument, see
\cite{odintsovfr,GRGrev}). In fact, General Relativity has been
successfully probed in the weak filed limit (e.g. Solar System
experiments) and also in this case there is room for alternative
theories of gravity which are not at all ruled out, as discussed
in several recent studies \cite{CST,AFRT,chiba,odisis}. In
particular, it is possible to show that several $f(R)$ models
could satisfy both cosmological and Solar System tests \cite{hu},
could be constrained as the scalar-tensor theories
\cite{CST,faulkner}, could give rise to new effects capable of
explaining the so called Pioneer anomaly (see for example
\cite{bertolami} and the references therein).

However, also considering the recent interest in the argument, a
comprehensive effective theory of gravity, acting consistently at
any scale, is far, up to now, to be found out, and this demands an
improvement of observational datasets and the search for
experimentally testable theories. A more pragmatic point of view
could be to ``reconstruct'' the suitable theory of gravity
starting from data. The main issues of this ``inverse '' approach
is matching consistently observations at different scales and
taking into account wide classes of gravitational theories where
``ad hoc'' hypotheses are avoided. In principle, the most popular
dark energy models can be achieved by considering  $f(R)$ theories
of gravity \cite{mimick,elizalde} and the same track can be
followed, at completely different scales, to match galactic
dynamics \cite{CCT}. This philosophy can be taken into account
also for the cosmological stochastic background of gravitational
waves (GW) which, together whit cosmic microwave background
radiation (CMBR) \cite{hwang}, would carry, if detected, a huge
amount of information on the early stages of the Universe
evolution. However, also in this case, a key role for the
production and the detection of this relic graviton background is
played by the adopted theory of gravity which gives rise to
specific early Universe models \cite{maggiore,babusci,corda}.

In this letter  we want to face the problem of how  a theory of
gravity with a Lagrangian of the form $R^{1+\epsilon}$, where
$\epsilon$ is a  deviation  (e.g. small) with respect to General
Relativity, could be related to the cosmological background of
GWs. They are  perturbations $h_{\mu\nu}$ of the metric
$g_{\mu\nu}$ which transform as  3-tensors. Following
\cite{weinberg}, the GW-equations in the transverse-traceless
gauge are
\begin{equation}
\square h_{i}^{j}=0\label{eq: 1}
\end{equation}
where
$\square\equiv(-g)^{-1/2}\partial_{\mu}(-g)^{1/2}g^{\mu\nu}\partial_{\nu}$
is the  d'Alembert operator;  these equations are deduced from the
weak-field limit, in vacuum, of the Einstein field equations.
 The Latin indexes run from 1
to 3, the Greek ones from 0 to 3. Our task is now to derive the
analog of Eqs.\ (\ref{eq: 1}) assuming a correction to the
Hilbert-Einstein action given by
\begin{equation}
\mathcal{A}=\frac{1}{2k}\int d^{4}x\sqrt{-g}f_0
R^{1+\epsilon}\,.\label{eq:2}
\end{equation}
 From now on we will assume units for which $f_0=1$. It is easy
to show that
\begin{equation}
R^{1+\epsilon}=R\cdot R^{\epsilon}\simeq R\left(1+(\log
R)\epsilon+{\cal O}(\epsilon^2)\right)\simeq R+\epsilon R\log R\,,
\end{equation}
and this form of the action can be assumed:  $i)$ to define the
right physical dimensions of the coupling constant; $ii)$ to
control the magnitude of the corrections with respect to the
standard Einstein gravity. Assuming a conformal transformation,
the extra degrees of freedom related to the higher order gravity
can be recast into a physically significant scalar field (for a
detailed discussion see \cite{ACCF,faraoni} and references
therein)
\begin{equation}
\widetilde{g}_{\mu\nu}=e^{2\Phi}g_{\mu\nu}\qquad \mbox{with}
\qquad e^{2\Phi}=\frac{df(R)}{dR}=f'(R)\simeq
(1+\epsilon+\epsilon\log R)\label{eq:3}
\end{equation}
where the prime indicates the derivative with respect to the Ricci
scalar $R$ and $\Phi$ is the ``conformal scalar field''. The
conformally equivalent Hilbert-Einstein action is
\begin{equation}
\mathcal{\widetilde{A}}=\frac{1}{2k}\int\sqrt{-\tilde{g}}d^{4}x\left[\widetilde{R}+
\mathcal{L}\left(\Phi\mbox{,}\Phi_{\mbox{;}\mu}\right)\right]\label{eq:4}\end{equation}
where $\mathcal{L}\left(\Phi\mbox{,}\Phi_{\mbox{;}\mu}\right)$ is
the conformal scalar field Lagrangian derived from
\begin{equation}
\widetilde{R}=e^{-2\Phi}\left(R-6\square\Phi-6\Phi_{;\delta}\Phi^{;\delta}\right)\,.\label{eq:6}\end{equation}
Deriving the field equations from (\ref{eq:4}), the GW-equations
are
\begin{equation}
\widetilde{\square}\tilde{h}_{i}^{j}=0\label{eq:7}
\end{equation}
expressed in the conformal metric $\widetilde{g}_{\mu\nu}$. Since
no scalar perturbation couples to the tensor part of gravitational
waves, we have
\begin{equation}
\widetilde{h}_{i}^{j}=\widetilde{g}^{lj}\delta\widetilde{g}_{il}=e^{-2\Phi}g^{lj}e^{2\Phi}\delta
g_{il}=h_{i}^{j}\label{eq:8}
\end{equation}
which means that $h_{i}^{j}$ is a conformal invariant.

As a consequence, the plane-wave amplitudes
$h_{i}^{j}(t)=h(t)e_{i}^{j}\exp(ik_{m}x^{m}),$ where $e_{i}^{j}$
is the polarization tensor, are the same in both metrics. This
fact will assume a key role in the following discussion. The
d'Alembert operator transforms as
\begin{equation}
\widetilde{\square}=e^{-2\Phi}\left(\square+2\Phi^{;\lambda}\partial_{;\lambda}\right)\label{eq:9}\end{equation}
and this means that the background is changing with frame while
the tensor plane-wave amplitudes not.

In order to study the gravitational waves in the cosmological
background, the operator (\ref{eq:9}) has to be specified for a
Friedmann-Robertson-Walker metric and then Eq.\ (\ref{eq:7})
becomes
\begin{equation}
\ddot{h}+\left(3H+2\dot{\Phi}\right)\dot{h}+k^{2}a^{-2}h=0\label{eq:10}
\end{equation}
being  $a(t)$ the scale factor,  $k$ the wave number and $h$
depending  only on the  surviving component of $h_{i}^{j}$ in the
homogeneous and isotropic metric.

\begin{figure}
\begin{tabular}{|c|c|}
\hline
\includegraphics[scale=0.6]{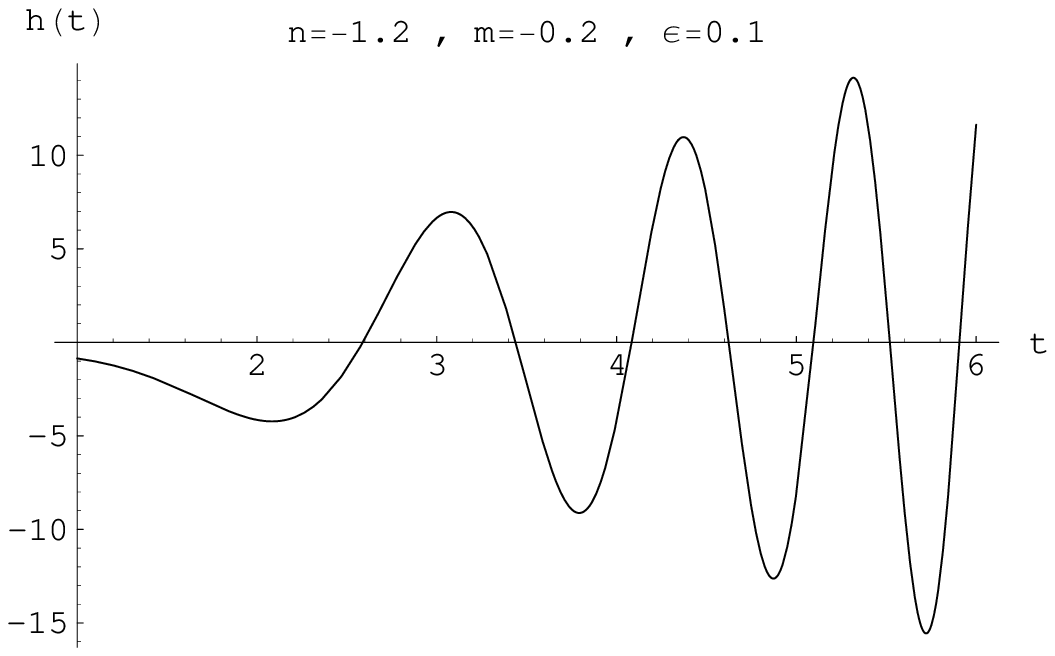}&
\includegraphics[scale=0.6]{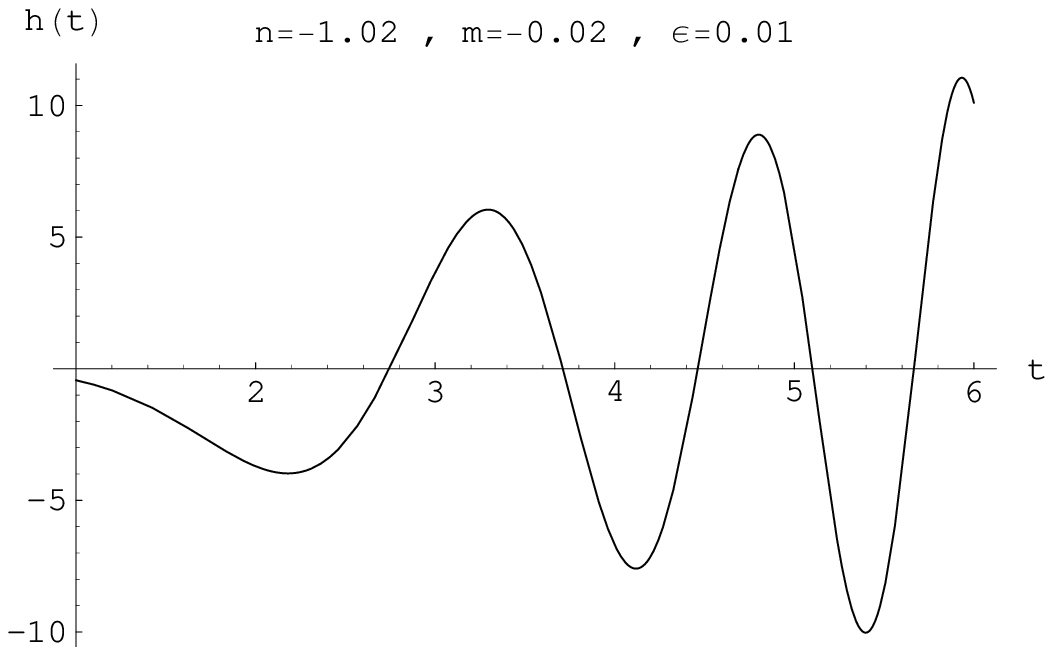}\tabularnewline
\hline
\includegraphics[scale=0.6]{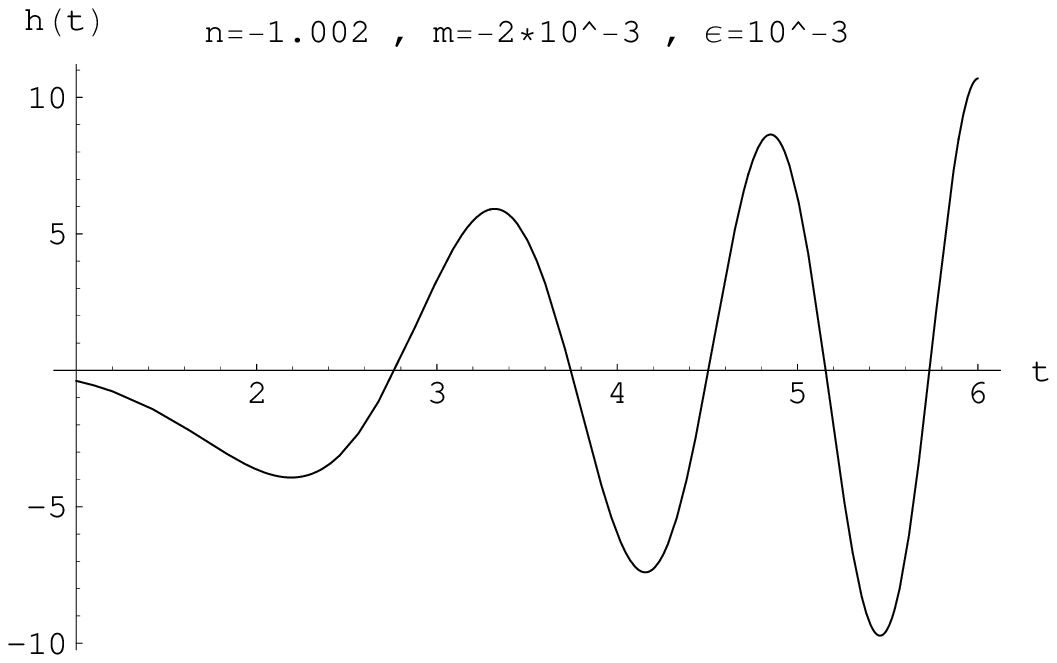}&
\includegraphics[scale=0.6]{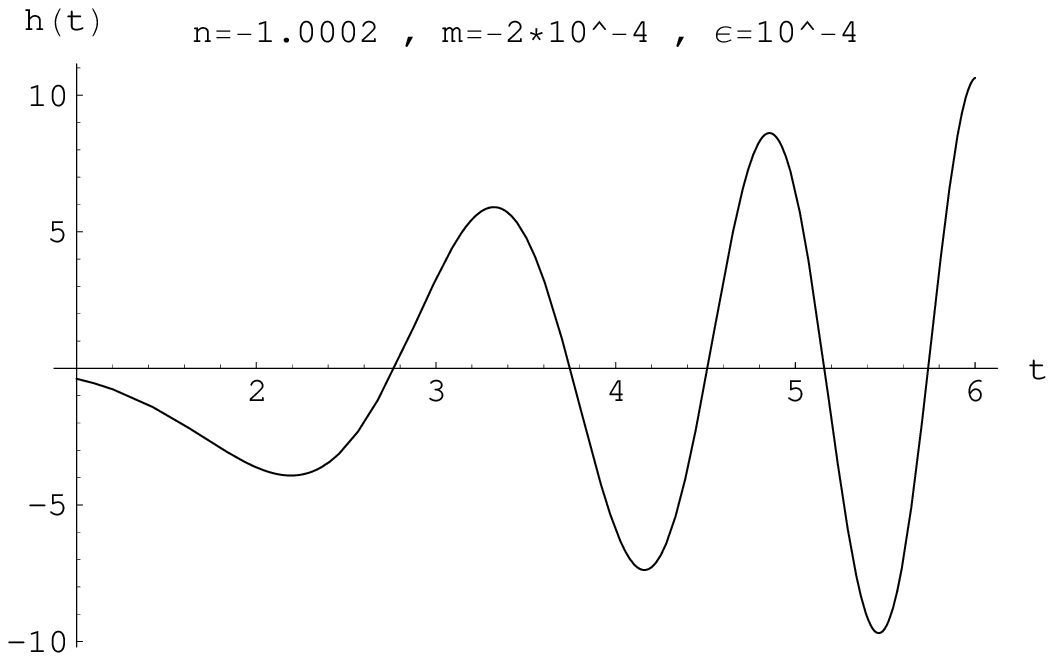}\tabularnewline
\hline
\includegraphics[scale=0.6]{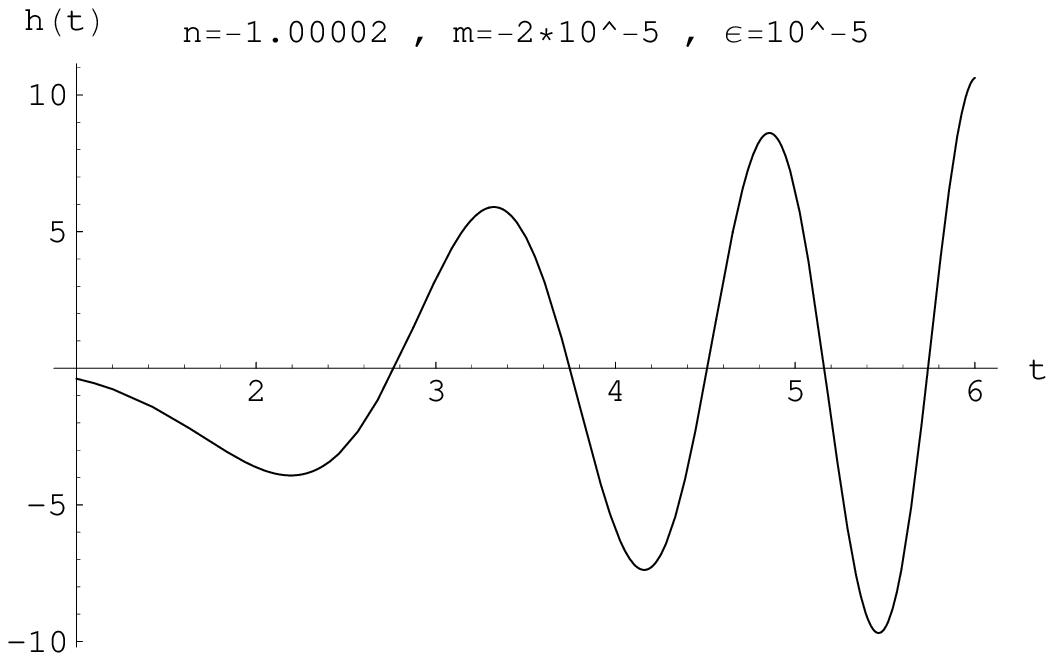}&
\includegraphics[scale=0.6]{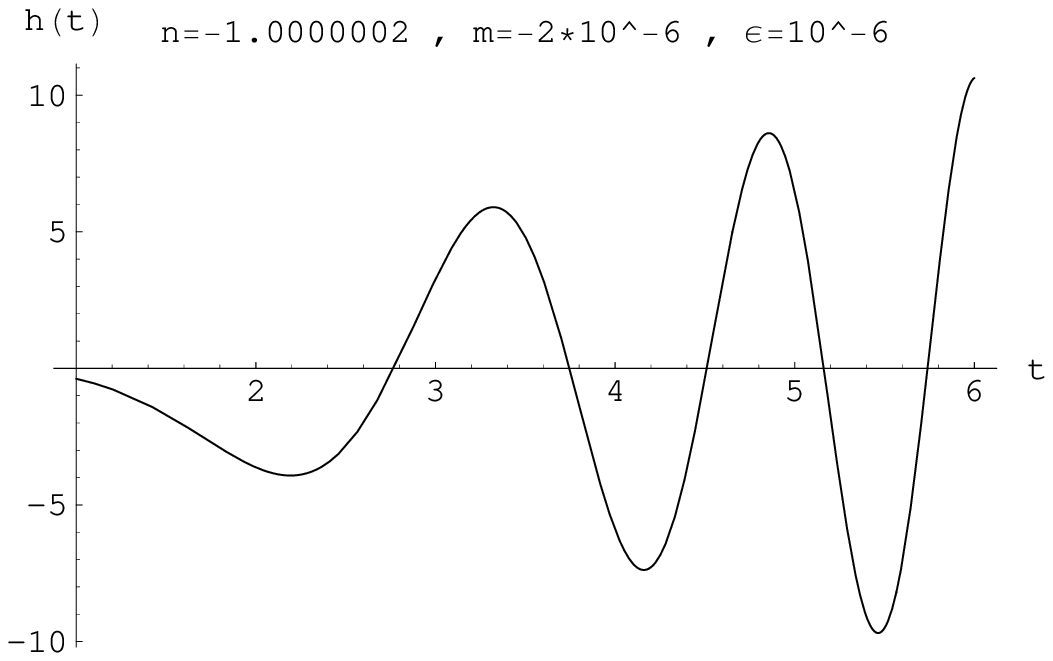}\tabularnewline
\hline
\includegraphics[scale=0.6]{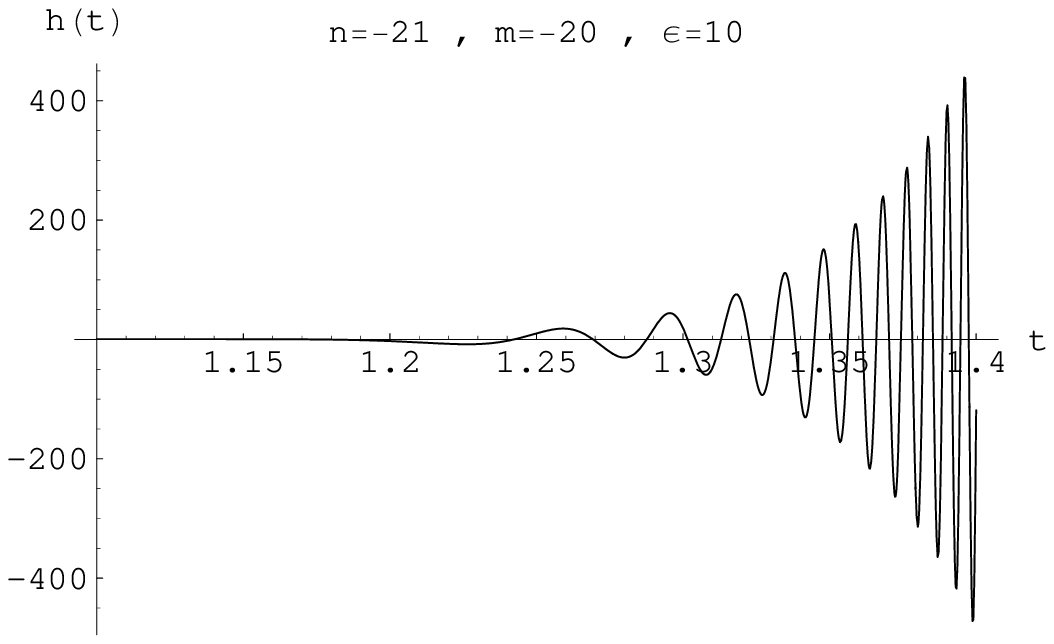}&
\includegraphics[scale=0.6]{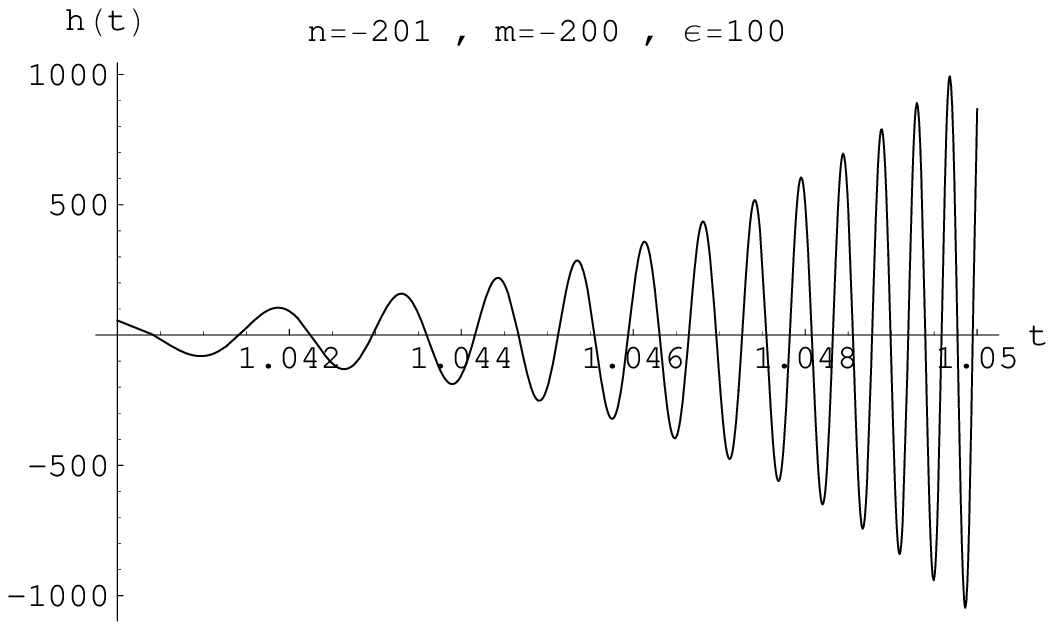}\tabularnewline
\hline
\end{tabular}
\caption {Qualitative evolution of the GW amplitude for some
choices of $a(t)$, $\Phi(t)$ and $R^{1+\epsilon}$. Time-scales,
amplitudes and "chirps" of GWs strictly depend on the value of
$\epsilon$ as it is especially shown in the last two (unrealistic)
cases ($\epsilon$ extremely large).} \label{fig:1}
\end{figure}

It is worth stressing that Eq.\ (\ref{eq:10}) applies to any
$f(R)$ theory whose conformal transformation can be defined as
${\displaystyle e^{2\Phi}=f'(R).}$ The GW amplitude depends on the
specific cosmological background $a(t)$ and the specific theory of
gravity which can be parameterized by $\epsilon$ in our  case
(\ref{eq:2}). For example, assuming time power law behaviors for
$a(t)$ and $f'(R(t))$, we have
\begin{equation}
f'(R)={f'}_{0}\left(t/t_{0}\right)^{m},\;\;
a(t)=a_{0}\left(t/t_{0}\right)^{n}\,.\label{eq:11}
\end{equation}
 For a generic $f(R)=R^{1+\epsilon}$, it is ${\displaystyle
\epsilon=-\frac{m}{2}}$. Eq.\ (\ref{eq:10}) can be recast in the
form
\begin{equation}
\ddot{h}+\left(3n-2\epsilon\right)t^{-1}\dot{h}+k^{2}a_{0}\left(t_{0}/t)\right)^{2n}h=0\label{eq:13}
\end{equation}
whose general solutions are
  $J_{\alpha}$'s ar Bessel functions. In Fig.\ (\ref{fig:1}) some examples
are given. The plots are labelled by the set of parameters $\{
m,\, n,\, \epsilon\}$ which assign the time evolution of $\Phi(t)$
and $a(t)$ with respect to a given power-law theory.

The time units are in terms of the Hubble radius $H^{-1}$. It is
clear that the conformally invariant plane-wave amplitude
evolution of the tensor GW strictly depends on the cosmological
background "tuned" by $\epsilon$.

Let us now take into account the issue of the production of GWs
contributing to the cosmological stochastic background. Several
mechanisms can be considered as vacuum fluctuations, phase
transitions \cite{maggiore},  cosmological populations of
astrophysical sources \cite{ferrari}  and so on. In principle, we
could seek for contributions due to every high-energy  process in
the early phases of the Universe evolution.

It is important to distinguish processes coming from transitions
like inflation, where the Hubble flow emerges in the radiation
dominated phase and processes, like the early star formation
rates, where the production takes place during  the dust dominated
era. In the first case, stochastic GW background is strictly
related to the cosmological model. This is the case we are
considering here which is strictly connected to the specific
theory of gravity. In particular, one can assume that the main
contribution to the stochastic background comes from the
amplification of vacuum fluctuations at the transition between an
inflationary phase and the radiation dominated era. However, in
any inflationary model, we can assume that the GWs generated as
zero-point fluctuations during the inflation undergo adiabatically
damped oscillations $(\sim 1/a)$ until they reach the Hubble
radius $H^{-1}$. This is the particle horizon for the growth of
perturbations. On the other hand, any  previous fluctuation is
smoothed away by the inflationary expansion. The GWs freeze out
for $a/k\gg H^{-1}$ and reenter the $H^{-1}$ radius after the
reheating in the Friedmann era (see also \cite{gri,allen}). The
reenter in the radiation-dominated or in the dust-dominated era
depends on the scale of the GW. After the reenter, GWs can be
detected by their Sachs-Wolfe effect on the temperature anisotropy
$\bigtriangleup T/T$ at the decoupling \cite{sachs}. If $\Phi$
acts as the inflaton  we have $\dot{\Phi}\ll H$ during the
inflation. Adopting the conformal time $d\eta=dt/a$, Eq.\
(\ref{eq:10}) reads
\begin{equation}
h''+2\frac{\chi'}{\chi}h'+k^{2}h=0\label{eq:16}
\end{equation}
where $\chi=ae^{\Phi}$. The derivation is  now with respect to
$\eta$. An inflationary behavior is achieved for
$a(t)=a_{0}\exp(Ht)$, then $\eta=\int dt/ a=(aH)^{-1}$ and
$\chi'/\chi=-\eta^{-1}$. The exact solution of (\ref{eq:16}) is
\begin{equation}
h(\eta)=k^{-3/2}\sqrt{2/k}\left[C_{1}\left(\sin k\eta-\cos k\eta\right)+C_{2}\left(\sin k\eta+\cos k\eta\right)\right]\label{eq:17}
\end{equation}
Inside the  radius $H^{-1}$, we have $k\eta\gg1.$ Considering the
absence of gravitons in the initial vacuum state, we have only
negative-frequency modes and then the adiabatic behavior is
\begin{equation}
h=k^{1/2}\sqrt{2/\pi}\frac{1}{aH}C\exp(-ik\eta)\,.\label{eq:18}
\end{equation}
where $C$ is the parameter to be tuned later. At the first horizon
crossing $(aH=k)$, the averaged amplitude
$A_{h}=(k/2\pi)^{3/2}\left|h\right|$ of the perturbation is
\begin{equation}
A_{h}=\frac{1}{2\pi^{2}}C\,.\label{eq:19}
\end{equation}
When the scale $a/k$ becomes larger than the Hubble radius
$H^{-1}$, the growing  mode of evolution is constant, i.e. it is
frozen. This situation corresponds to the limit $-k\eta\ll 1$ in
Eq.\ (\ref{eq:17}). Since $\Phi$ acts as the inflaton field, we
have $\Phi\sim 0$ at reenter (after the end of inflation). Then
the amplitude $A_{h}$ of the wave is preserved until the second
horizon crossing; after  it can be observed, in principle, as an
anisotropy perturbation on the CMBR. It can be shown that
$\bigtriangleup T/T\lesssim A_{h}$, as an upper limit to $A_{h}$,
since other effects can contribute to the background anisotropy.
From this consideration, it is clear that the only relevant
quantity is the initial amplitude $C$ in Eq.\ (\ref{eq:18}), which
is conserved until the reenter. Such an amplitude depends  on the
fundamental mechanism generating perturbations. Inflation gives
rise to processes capable of producing perturbations as zero-point
energy fluctuations. Such a mechanism depends on the gravitational
interaction and then $(\bigtriangleup T/T)$ could constitute a
further constraint to select a suitable theory of gravity.
Considering a single graviton in the form of a monochromatic wave,
its zero-point amplitude is derived through the commutation
relations:
\begin{equation}
\left[h(t,x),\,\pi_{h}(t,y)\right]=i\delta^{3}(x-y)\label{eq:20}
\end{equation}
calculated at a fixed time $t$, where the amplitude $h$ is the
field and $\pi_{h}$ is the conjugate momentum operator. Writing
the Lagrangian for $h$
\begin{equation}
\widetilde{\mathcal{L}}=\frac{1}{2}\sqrt{-\widetilde{g}}\widetilde{g}^{\mu\nu}h_{;\mu}h{}_{;\nu}\label{eq:21}
\end{equation}
in the conformal FRW metric $\widetilde{g}_{\mu\nu}$  (recall that
the amplitude $h$ is conformally invariant), we obtain
\begin{equation}
\pi_{h}=\frac{\partial\widetilde{\mathcal{L}}}{\partial\dot{h}}=e^{2\Phi}a^{3}\dot{h}\label{eq:22}\end{equation}

Eq.\ (\ref{eq:20}) becomes
\begin{equation}
\left[h(t,x),\,\dot{h}(y,y)\right]=i\frac{\delta^{3}(x-y)}{a^{3}e^{2\Phi}}\label{eq:23}
\end{equation}
and the fields $h$ and $\dot{h}$ can be expanded in terms of
creation and annihilation operators
\begin{equation}
h(t,x)=\frac{1}{(2\pi)^{3/2}}\int d^{3}k\left[h(t)e^{-ikx}+h^{*}(t)e^{+ikx}\right],\label{eq:24}
\end{equation}
\begin{equation}
\dot{h}(t,x)=\frac{1}{(2\pi)^{3/2}}\int d^{3}k\left[\dot{h}(t)e^{-ikx}+\dot{h}^{*}(t)e^{+ikx}\right].\label{eq:25}
\end{equation}

The commutation relations in conformal time are then
\begin{equation}
\left[hh'^{*}-h^{*}h'\right]=\frac{i(2\pi)^{3}}{a^{3}e^{2\Phi}}\label{eq:26}
\end{equation}
Using (\ref{eq:18}) and (\ref{eq:19}), we obtain
$C=\sqrt{2}\pi^{2}He^{-\Phi}$, where $H$ and $\Phi$ are calculated
at the first horizon-crossing and
\begin{equation}
A_{h}=\frac{\sqrt{2}}{2}He^{-\Phi}\simeq\frac{\sqrt{2}}{2}H\left(1-\Phi+\frac{1}{2}\Phi^2\cdot\cdot\cdot\right)
=\frac{\sqrt{2}}{2}H\left(1-\frac{\epsilon}{2}\log R +
\frac{\epsilon^2}{8}(\log R)^2\cdot\cdot\cdot \right)
\,\,,\label{eq:27}
\end{equation}
or, in general,
\begin{equation} A_{h}=\frac{H}{\sqrt{2f'(R)}}\,.\end{equation}
Clearly the amplitude of GWs produced during inflation depends on
the given theory of gravity that, if different from General
Relativity, gives extra degrees of freedom capable of affecting
the cosmological dynamics. On the other hand, the Sachs-Wolfe
effect related to the CMBR temperature anisotropy could constitute
a test for the theory of gravity at early epochs, i.e. at very
high redshift. This probe could give further constraints on the
GW-stochastic background, if Extended Theories of Gravity are
independently probed at other scales. In summary, assuming that
primordial vacuum fluctuations produce GWs, beside scalar
perturbations, kinematical distortions and so on, the initial
amplitude of these ones is a function of the assumed theory of
gravity and then the stochastic background can, in a certain
sense, be ``tuned'' by the theory. Conversely, data coming from
the Sachs-Wolfe effect could contribute to select a suitable
$\epsilon$  which can be consistently  matched with other
observations.  If eventually, one finds  $\epsilon$ rigorously
zero, this could be a further test for General Relativity also at
very early cosmological scales. This result could be obtained very
soon through space and ground based interferometers
\cite{pramana}.

\end{document}